\documentclass{jetpl}
\twocolumn

\title{Constraints on the Number Density of Evaporating Primordial
Black Holes for the Chromospheric Evaporation Models}

\rtitle{Constraints on the Number Density...}

\sodtitle{Constraints on the Number Density of Evaporating
Primordial Black Holes for the Chromospheric Evaporation Models}

\author{V.\,B.\,Petkov, E.\,V.\,Bugaev, P.\,A.\,Klimai, D.\,V.\,Smirnov}

\rauthor{PETKOV et al}

\sodauthor{PETKOV et al}

\address{Institute for Nuclear Research, Russian Academy of Sciences, Moscow, 117312 Russia\\
e-mail: pklimai@gmail.com}

\dates{December 6, 2007} {*}

\abstract{ Cosmic gamma-ray bursts with primary-photon energies
$\ge 10$ GeV are sought in the data from the Andyrchy array
obtained in the mode of detection of a single cosmic-ray component
during a net observation period of 2005.4 d. The distribution of
fluctuations of the detector counting rate agrees with the
expected cosmic-ray background, the only exception being an event
with a deviation of $7.9\sigma$. Constraints on the number density
of evaporating primordial black holes in a local region of the
Galaxy are obtained for the chromospheric evaporation models. }


\begin{document}
\maketitle

\section*{INTRODUCTION}

Primordial black holes (PBHs) can be formed in the early Universe
due to the gravitational collapse of primordial cosmological
density fluctuations. Theoretical predictions of the probability
of PBH formation are strongly dependent on the accepted theory of
gravitation and the model of gravitational collapse. The process
of black-hole evaporation, which is used to seek them
experimentally, is also far from being completely understood.
Thus, PBH detection will provide important information on the
early Universe and can be a unique test of the general theory of
relativity, cosmology, and quantum gravity \cite{r1}. Knowledge of
the PBH spatial distribution is important for their direct search.
The number density of PBHs in our Galaxy can be many orders of
magnitude higher than the average number density of PBHs in the
Universe \cite{r2}, so that the constraints based on the results
of direct searches can be much more stringent than those implied
by measuring the diffuse extra-Galactic $\gamma$-ray background.

\begin{figure}[b]
\center{\includegraphics[width=\columnwidth]{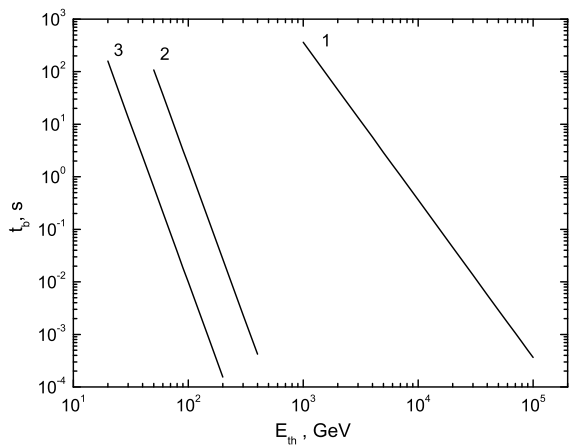}}
\caption{{\bf Fig.1.} Burst duration versus the threshold energy
of gamma-ray photons for various evaporation models: (1) the model
without a chromosphere \cite{r4}, (2) the DK02 model with a
chromosphere \cite{r5}, and (3) the H97 model with a chromosphere
\cite{r6}.}
\end{figure}

The bursts of high-energy $\gamma$ rays are generated at the final
stages of PBH evaporation. Since the calculated temporal and
energy characteristics of such bursts depend on the accepted
theoretical evaporation model \cite{r3}, the experimental
technique used to detect PBH emission and, correspondingly, the
resulting constraints on the number density of PBHs in the local
Universe are strongly model dependent. In the model without a
chromosphere \cite{r4}, the evaporating particles do not interact,
all quarks propagate freely, and evaporating particles fragment
independently of each other. The photon spectrum is formed by the
quark fragmentation and decay of unstable hadrons; hence, this
spectrum is not thermal. According to the DK02 \cite{r5} and H97
\cite{r6} chromospheric models, the interacting evaporated
particles form a quasi-chromosphere, which results in the strong
energy fragmentation and, accordingly, in the steep photon
spectrum at high energies. The spectra of $\gamma$ rays emitted by
PBHs depend on the time until the end of the black-hole
evaporation. The evaporation models were analyzed in more detail
in \cite{r3}. The time interval until the end of black-hole
evaporation during which 99\% of the $\gamma$-ray photons with
energies $E_\gamma \ge E_{\rm th}$ are emitted is called the burst
duration for the threshold $E_{\rm th}$. The burst duration as a
function of the threshold energy is plotted in Fig.1 for three
evaporation models. Until recently, the bursts of high-energy
$\gamma$ rays from the final stage of PBH evaporation were sought
using several arrays for detecting the extensive air showers
(EASs) from cosmic rays \cite{r7} and the Whipple Cherenkov
telescope \cite{r8}. Since the threshold energy of the primary
$\gamma$-ray photons is high and the duration of the high-energy
$\gamma$-ray burst predicted by the chromospheric models is too
short (much less than the dead time of these arrays), the results
of these experiments can be interpreted only within the framework
of the evaporation model without a chromosphere.

\begin{figure}[t]
\center{\includegraphics[width=\columnwidth]{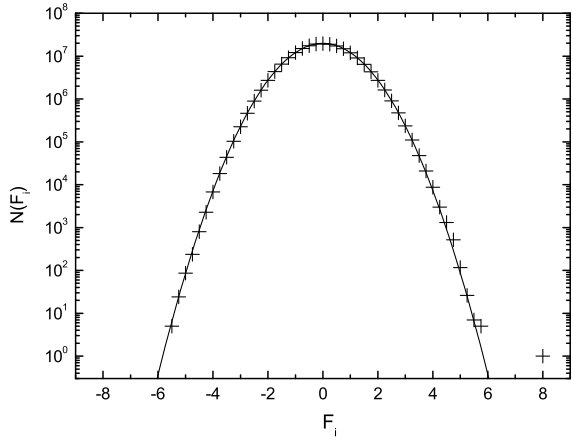}}
\caption{{\bf Fig. 2.} Distribution in the parameter $F_i$ for the
data gathered during nine years; the Gaussian fit is shown by the
solid line.}
\end{figure}

Within the framework of the chromospheric models, the PBH events
can be sought directly by the method of detecting spikes in the
total counting rate of the arrays operating in the regime of the
detection of a single cosmic-ray component. The particle arrival
direction in this regime is not determined, while the effective
energy of the primary gamma-ray photons depends mainly on the
altitude of the array above sea level. Such a technique was
employed earlier to seek cosmic gamma-ray bursts with energies
exceeding a few GeVs \cite{r9,r10,r11}.

\section*{EXPERIMENT}

The Andyrchy EAS array of the Baksan Neutrino Observatory
(Institute for Nuclear Research, Russian Academy of Sciences) is
located at an altitude of 2060m above sea level and consists of 37
scintillation detectors. The area of each detector is 1 m$^2$. To
detect a single cosmic-ray component, the total counting rate of
all of the detectors is measured each second. The search for
$\gamma$-ray bursts using this technique is carried out against
the high cosmic-ray background ($\bar \omega = 11440 \; {\rm s} ^
{-1} $), which requires the highly stable and reliable performance
of all equipment. The monitoring is realized through simultaneous
measurements (with 1-s acquisition rate) of the counting rates in
the four parts of the array comprising 10, 9, 9, and 9 detectors,
respectively. A detailed description of the array and its
operating parameters is given in \cite{r12}. The probabilities
$P(E_\gamma, \theta)$ of the detection of the secondary particles
that are created by the primary $\gamma$-ray photons with energy
$E_\gamma$ and that incident onto the array at the zenith angle
$\theta$ in a hypothetical infinite-area Andyrchy detector were
obtained by simulating the electromagnetic cascades in the
atmosphere and detector \cite{r11}. Since the probability of the
detection of a $\gamma$-ray photon is a fairly smooth function of
the photon energy, the median energy of the primary $\gamma$-ray
photons detected by the detector depends on their energy spectrum.
In the case of a source with zero zenith angle having a power-law
spectrum with an exponent of $-2.0$, the median energy of the
primary  $\gamma$-ray photons is equal to 10 GeV.

Deviations in the array counting rate lasting $\Delta t \le 1 $s
are sought using the parameter $F_i$ that is equal to the
deviation (measured in units of the Poisson sigma) of the number
of counts $k_i$ during the {\it i}th second of a 15-min interval
from the average number of counts $\bar k$ during this interval:
$F_i = (k_i - \bar k)/ \sqrt{\bar k}$. Since variations in the
cosmic-ray intensity over a time of 15 min are negligible in the
first approximation and the average counting rate is fairly high,
one can expect that the parameter $F_i$ has a Gaussian
distribution with the zero mean value $V = 0$ and unit standard
deviation $\sigma = 1.0$. The parameter $D_i$ is used to
characterize the differences in the counting rates of the array
parts: $D_i = \frac{1}{4} \sum_{j=1}^{4} (F_i^j - \bar F_i)^2 $.
Here, $F_i^j$ is the deviation of the {\it j}th part of the array
and $\bar F_i$ is the average of the four values for the {\it i}th
second of a 15-min interval. Further processing was done only for
the 1-s points for which the condition $D_{i} \le D_{\rm
bound}(F_i)$ was valid. The values $D_{\rm bound}(F_i)$ were
obtained by the Monte Carlo method under the assumption that $k_i$
has a Poisson distribution. This condition makes it possible to
reject points for which the differences in the counting rates of
the array parts are unreasonably large, i.e., eliminate the
instrumental error. The useful events (1-s points) can be rejected
with a probability of $2 \times 10^{-9} $ for all of the events
and with a probability of $1.3 \times 10^{-3}$ for events with
$F_i \ge 5$. About 0.01\% of events were rejected from the entire
volume of the experimental data according to this criterion.

Figure 2 shows the experimental distribution in the parameter
$F_i$ plotted using the data gathered during nine years (2005.4 d
of the net observation time). The experimental data up to $ \sim 6
\sigma$ are fitted well by a Gaussian distribution with the mean
value $V = -(0.0025 \pm 0.0004)$ and standard deviation $\sigma =
1.006$. The only event with a large ($7.9 \sigma$) deviation was
detected April 17, 2002 at 17:31:29 UT. No $\gamma$-ray bursts
were detected at this time by space-borne instruments.

\section*{EXPECTED SIGNAL FROM PBHS}

Let a PBH be located at distance $r$ from the array and observed
by this array in the direction specified by the zenith angle
$\theta$. Then, the mean number of $\gamma$-ray photons detected
by the array is equal to
\begin{equation}
\bar n(\theta, r) = \frac{S(\theta)}{4\pi
r^2}\,N(\Delta t,\theta) ,
\end{equation}
where $N(\Delta t,\theta)$ is the total number of the gamma-ray
photons emitted by the PBH that can be detected by the array:
\begin{equation}
N(\Delta t,\theta) = \int\limits_{0}^{\infty}dE_\gamma \,
P(E_{\gamma},{\theta})\,\frac{dN_{\gamma}}{dE_{\gamma}}(\Delta t).
\end{equation}
Here, $(dN_{\gamma}/dE_{\gamma})(\Delta t)$ is the spectrum of
gamma-ray photons emitted by the PBH during the time interval
$\Delta t$ until the end of the PBH evaporation and $S(\theta)$ is
the area of the array. The number of bursts detected during the
entire observation time $T$ can be written as
\begin{equation}
N=\rho_{\rm pbh}\,T\,V_{\rm eff} ,
\end{equation}
where
\begin{equation}
V_{\rm eff}=\int d \Omega {\int \limits_{0}^{\infty} dr r^2
F(n,\bar n(\theta, r))}
\end{equation}
is the effective volume of the space surveyed by the array,
$\rho_{\rm pbh}$ is the number density of the evaporating PBHs,
and $F(n, \bar n) = e^{- \bar n} \bar n ^ n /n!$ is the Poisson
probability of the detection of $n$ events for the mean value
$\bar n$.

\section*{CONSTRAINTS ON THE PBH NUMBER
DENSITY FOR THE CHROMOSPHERIC MODELS}

The upper limit for the PBH number density (excluding the single
spike in the counting rate) can be obtained from the absence of
events with deviations $\ge 6\sigma$ for $ \Delta t = 1 $s, i.e.,
the effective volume of space (4) surveyed by the array is
calculated for $n = 6\sigma = 642$. If the evaporating PBHs are
uniformly distributed over the local Galactic region, then the
upper limit $\rho_{\rm lim}$ for the number density of evaporating
PBHs at the 99\% confidence level is calculated by the formula
\begin{equation}
\rho_{\rm lim} = \frac{4.6}{V_{\rm eff}\,T} ,
\end{equation}
where the net observation period is $T = 5.5 $yr. The substitution
of the $V_{\rm eff}$ values for each chromospheric evaporation
model into Eq. (5) yields a 99\%-C.L., with the upper limits $1.8
\times 10^{12}$ and $1.7 \times 10^{13} \; \rm pc^{ - 3 } yr^{-1}$
for the DK02 and H97 evaporation models, respectively.

\section*{PROBABLE PBH EVENT}

Let us assume that the event of 7.9-$\sigma$ deviation is caused
by the $\gamma$-ray burst from an evaporating PBH. Then, according
to the evaporation model \cite{r4} without a chromosphere, the
average number density of evaporating PBHs is equal to $2.9 \times
10^{12} \; \rm pc^{-3} yr^{-1}$, which is many orders of magnitude
higher than the upper limits obtained earlier in several
experiments \cite{r7,r8}.

According to the chromospheric models, the average number density
of PBHs is equal to $\bar \rho = 7.8 \times 10^{11}$ and $7.2
\times 10^{12} \; \rm pc^{-3} yr^{-1}$ for the DK02 and H97
models, respectively. Note that a single spike in the total
counting rate was also observed at the EAS-TOP array \cite{r9}.
The deviation amounted to 20.6$\sigma$ for $\Delta t$ = 2 s. If
this spike is a PBH event, then the average number density of
evaporating PBHs amounts to $\bar \rho = 6.2 \times 10^{12}$ and
$3.7 \times 10^{13} \; \rm pc^{-3} yr^{-1}$ for the DK02 and H97
models, respectively. Despite the significant (more than an order
of magnitude) difference between the mean values, the 99\%
confidence intervals for each model intersect with each other.
Thus, if the events with large deviations in the total counting
rates of the EAS-TOP and Andyrchy arrays result from $\gamma$-ray
bursts from evaporating PBHs, then the average number density of
the evaporating PBHs is $7.5 \times 10^{10} - 5.2 \times 10^{12}$
and $4.5 \times 10^{11} - 4.8 \times 10^{13} \; \rm pc^{-3}
yr^{-1}$ for the DK02 and H97 models, respectively.

\vspace{0.5cm}

This work was supported by the Russian Foundation for Basic
Research (project no. 06-02-16135), the Presidium of the Russian
Academy of Sciences (Basic Research Program "Neutrino Physics"),
and the Council of the President of the Russian Federation for
Support of Young Scientists and Leading Scientific Schools
(project no. NSh-4580.2006.02).

\vspace{0.5cm}
{\sl Translated by A.V. Serber}.

\end{document}